\documentstyle[12pt,epsfig]{article}
\begin{document}
\setlength{\baselineskip}{14pt}
\parindent=40 mm
\title{\Large\bf Chiral symmetry restoration 
and parity mixing}
\author{ G. Chanfray, J. Delorme and M. Ericson\thanks{ Also at
 Theory Division, CERN,
CH-1211 Geneva
 23,
Switzerland} \\
{\small Institut de Physique Nucl\'{e}aire et IN2P3, CNRS,
Universit\'{e} Claude Bernard Lyon I,}
{\protect \vspace{-2mm}}
\and {\small 43 Bd. du 11 Novembre,
F-69622 Villeurbanne Cedex, France} \\ }

\date{}

\maketitle

\begin{abstract}
We derive the expressions of the vector and axial current from a chiral
Lagrangian restricted to nucleons and pions. They display mixing terms
 between the axial and vector currents. We study the
modifications in the nuclear medium of the coupling constants of the axial
current, namely the pion decay constant and the nucleonic axial one
 due to the requirements of chiral symmetry. We express the
renormalizations in terms of the local scalar pion density.
 The latter also governs the quark condensate evolution and we
discuss the link between this evolution and the renormalizations. In the case
of the nucleon axial coupling constant this renormalization
corresponds to a new type of pion exchange currents, with two exchanged
pions. We give an estimate for the resulting quenching. Although
moderate it helps explaining the quenching experimentally observed.

\end{abstract}

\centerline{PACS numbers: 12.39.Fe 24.85.+p 23.40.Bw}
\vskip 4mm
\parindent = 30pt

\section*{Introduction}
The problem of the restoration in dense or hot matter of the chiral
symmetry of the strong interactions, which is spontaneously violated in
the QCD vacuum has been extensively addressed. The interest has largely
focused on the quark condensate, considered as the order parameter.
For independent particles the evolution of the quark condensate with
density or temperature is governed by the sigma commutator of the
particles present in the system with  the simple following expression:
\begin{equation}
\frac{<\overline{q} q(\rho)>}{<\overline{q} q(0)>} = 1 -
\sum_n\frac{\rho_n^s\;\Sigma_n}{f_{\pi}^2m_{\pi}^2}
\label{sigma}
\end{equation}
where the sum extends over the species present in the medium, $\rho^s$ is
their scalar density and $\Sigma$ their sigma commutator.
 Pions play a crucial role in the restoration process especially in the heat
bath where they enter as the lightest particles created
by the thermal fluctuations. In the nuclear medium the main ingredients are the
nucleons, with some corrections from the exchanged pions. At normal nuclear
 density the magnitude of the condensate has dropped by
about 1/3, a large amount of restoration. It is essentially the effect of
the nucleons adding their effects independently, the corrections due to
the interaction being small.
Such a large amount of restoration raises the question about
manifestations directly linked to the symmetry. If there is no
spontaneous violation of the symmetry, {\it i.e.}, if it is realized in the
Wigner mode, the hadron masses vanish  or there 
exist parity doublets, each hadronic state being degenerate with its
chiral partner. It is therefore legitimate to believe that the large
amount of restoration at normal density manifests itself either by a decrease
 of the hadron masses, or by a mixing between opposite parities.  A link
between the evolution of the hadron masses and the amount of restoration
has been suggested~\cite{BR}. But it cannot be a straightforward one. Indeed
 the density or temperature evolution of the masses
cannot have a direct relation to that of the condensate
, as follows from the works of several authors~\cite{LS,EI,BIR}.
 On the other hand
the significance of chiral symmetry restoration for the parity mixing  was
first established by Dey et al.~\cite{IOF} for the thermal case. They showed 
that in a pion gas a mixing occurs between the vector and axial
correlators. It arises from  the
emission or absorption of s-wave thermal pions, which changes the parity
of the system. The mixing goes along with a quenching effect of the correlators,
 which, to first order in the pion density,  
equals 4/3 of the quenching of the quark condensate. These points were also
 made by Steele et al.~\cite{ZAH}. The extension
 of the formalism of Dey et al.
to finite densities has been attempted by  Krippa~\cite{KRI}.

The aim of this work is the discussion of the implications of chiral symmetry
restoration in the nuclear medium, in
a world restricted here to nucleons and pions. The only transitions
allowed in the nucleus are then nuclear transitions or pion production. We
give the explicit expressions of the axial and vector
current in a formalism based on chiral lagrangians. We will show that the
 nuclear pions renormalize the coupling
 constants of the axial current and that the renormalization
can be expressed in terms of the pion scalar density. This last quantity also
enters in the quark condensate evolution. However the
complexity of the nuclear interactions bars a simple link between this
 evolution, which is an average concept, and the
renormalizations.  For instance, for the axial coupling constant $g_A$
 the detailed spatial structure of the pion scalar density is needed. 

Our article is organized as follows. In section 1 we derive the expressions
of the axial and vector currents from the chiral lagrangians. In section 2
we use these expressions to study the renormalization of the pion decay
constant in the hot pion gas and in the nuclear medium. The thermal case is
only introduced as an illustration of the method since the results are already
known. In section 3 we apply the same technique to the axial coupling constant.
 To account for the nucleon-nucleon correlations we express the renormalization
in the traditional treatment by the meson exchange currents. We give an estimate
of the quenching of the axial coupling constant. We also discuss the
renormalization of the Kroll-Ruderman matrix element of pion photoproduction. 

\bigskip
\section{The Lagrangian and the currents} 
Our starting point is the chiral Lagrangian in the form introduced by
Weinberg. We use, as in our previous work of ref.~\cite{DCE}, the version of
 Lynn~\cite{LYN}, which allows one to obtain the nucleon sigma commutator
 in the tree approximation. The Lagrangian writes:
\begin{eqnarray}
{\cal L}& = & -\frac{1}{2} m_{\pi}^2 \frac{\displaystyle{\hbox{\boldmath$
\phi $\unboldmath}}^2}{\displaystyle 1 + {\hbox{\boldmath$\phi $\unboldmath}}
 ^2/4f_{\pi}^2} + \frac{1}{2}\frac{\displaystyle \partial_{\mu}
{\hbox{\boldmath$\phi $\unboldmath}}\cdot 
\partial^{\mu}{\hbox{\boldmath$\phi $\unboldmath}}}
{\displaystyle(1 + {\hbox{\boldmath$\phi $\unboldmath}}^2/4f_{\pi}^2)^2}
 \nonumber  \\
& & + 2\sigma_N \overline{\psi}\psi
\frac{\displaystyle{\hbox{\boldmath$\phi $\unboldmath}}^2/4f_{\pi}^2}
{\displaystyle 1 + {\hbox{\boldmath$\phi $\unboldmath}}^2/4f_{\pi}^2} +
\overline{\psi}(i\gamma_{\mu}\partial^{\mu}-M)\psi \nonumber \\
& & - \frac{1}{4f_{\pi}^2}
\frac{\displaystyle\overline{\psi}\gamma_{\mu}{\hbox{\boldmath$(\tau \times
\phi)$\unboldmath}} \cdot \partial^{\mu}{\hbox{\boldmath$\phi $\unboldmath}}
\psi}{\displaystyle 1 +
{\hbox{\boldmath$\phi $\unboldmath}}^2/4f_{\pi}^2}
+
\frac{g_A}{2f_{\pi}}
\frac{\displaystyle\overline{\psi}\gamma_{\mu}\gamma_5
{\hbox{\boldmath$\tau $\unboldmath}}\cdot\partial^{\mu}
{\hbox{\boldmath$\phi $\unboldmath}}\psi}{\displaystyle 1
+{\hbox{\boldmath$\phi $\unboldmath}}^2/4f_{\pi}^2} \; .
\label{lag}
\end{eqnarray}

 We have to specify the quantity $\sigma_N$
associated with the nucleon density in eq.~(\ref{lag}). The free nucleon
sigma commutator $\Sigma_N$ cannot be entirely attributed to the pion cloud.
We define $\sigma_N$ to be the difference between the total and pionic 
contributions:
\begin{equation}
\Sigma_N = \sigma_N + \frac{1}{2} m_{\pi}^2 \int
 d{\hbox{\boldmath$x $\unboldmath}}\langle N \vert 
{\hbox{\boldmath$\phi^2(x) $\unboldmath}}\vert N \rangle .
\label{sig}
\end{equation}

For instance in a description of the nucleon in terms of
valence quarks and pions, the pionic contribution is approximatively
1/2 to 2/3 of the total value~\cite{JCT,BMG}. 

From the Lagrangian of eq.~(\ref{lag}) we derive the expressions of the axial
 and isovector vector currents:
\begin{eqnarray}
{\hbox{\boldmath${\cal A} $\unboldmath}}_{\mu}  & = & 
f_{\pi}\frac
{\displaystyle\partial_{\mu}{\hbox{\boldmath$\phi $\unboldmath}}}
{\displaystyle 1 +
{\hbox{\boldmath$\phi $\unboldmath}}^2/4f_{\pi}^2}
-\frac{1}{2f_{\pi}}\frac{\displaystyle\big[({\hbox{\boldmath$\phi
 \times $\unboldmath}}\partial_{\mu}{\hbox{\boldmath$\phi)\times\phi
 $\unboldmath}}\big]}{\displaystyle (1 +
{\hbox{\boldmath$\phi $\unboldmath}}^2/4f_{\pi}^2)^2} \nonumber \\ 
   &  & + \frac{g_A}{2}\overline{\psi}\gamma_{\mu}\gamma_5
{\hbox{\boldmath$\tau $\unboldmath}}\psi
+\frac{g_A}{4f_{\pi}^2}\frac{\displaystyle\overline{\psi}\gamma_{\mu}\gamma_5
\big[{\hbox{\boldmath$(\tau\times\phi)\times\phi $\unboldmath}}\big]\psi}
{\displaystyle 1 +
{\hbox{\boldmath$\phi $\unboldmath}}^2/4f_{\pi}^2}
-\frac{1}{2f_{\pi}}\frac{\displaystyle\overline{\psi}\gamma_{\mu}
{\hbox{\boldmath$(\tau\times\phi) $\unboldmath}}\psi}
{\displaystyle 1 +
{\hbox{\boldmath$\phi $\unboldmath}}^2/4f_{\pi}^2}
\label{acur}   \\
& & \nonumber \\
{\hbox{\boldmath${\cal V} $\unboldmath}}_{\mu} & = &
\frac{\displaystyle({\hbox{\boldmath$\phi
 \times $\unboldmath}}\partial_{\mu}{\hbox{\boldmath$\phi)
 $\unboldmath}}}{\displaystyle( 1 +
{\hbox{\boldmath$\phi $\unboldmath}}^2/4f_{\pi}^2)^2} \nonumber \\
& & +\frac{1}{2}\overline{\psi}\gamma_{\mu}
{\hbox{\boldmath$\tau $\unboldmath}}\psi
+\frac{1}{4f_{\pi}^2}\frac{\displaystyle\overline{\psi}\gamma_{\mu}
\big[{\hbox{\boldmath$(\tau\times\phi)\times\phi $\unboldmath}}\big]\psi}
{\displaystyle 1 +
{\hbox{\boldmath$\phi $\unboldmath}}^2/4f_{\pi}^2}
-\frac{g_A}{2f_{\pi}}\frac{\displaystyle\overline{\psi}\gamma_{\mu}\gamma_5
{\hbox{\boldmath$(\tau\times\phi) $\unboldmath}}\psi}
{\displaystyle 1 +
{\hbox{\boldmath$\phi $\unboldmath}}^2/4f_{\pi}^2}\; .
\label{vcur}
\end{eqnarray}

The conservation law of the vector current can be shown, using the equations of
motion for the nucleon and the pion fields.
The divergence of the axial current instead  satisfies the following
relation:
\begin{equation}
\partial^{\mu}{\hbox{\boldmath${\cal A}$\unboldmath}}_{\mu} = -
f_{\pi}m_{\pi}^2\;\frac{{\hbox{\boldmath$\phi$\unboldmath}}}
{\displaystyle 1 + {\hbox{\boldmath$\phi$\unboldmath}}^2/4f_{\pi}^2}
\;(1-\sigma_N\frac{\overline{\psi}\psi}{f_{\pi}^2m_{\pi}^2}) \; .
\label{div}
\end{equation}

Some comments on the expressions~(\ref{acur}) and~(\ref{vcur}) are in order.
 Let us first discuss the free
case.  We recognize in some of the terms the usual expressions for the
vector or axial current coupled to a free nucleon or pion. In addition the axial
current can create one or more pions, either in free space (first
terms of eq.~(\ref{acur})), or when it acts on the nucleon 
via a term (last one of eq.~(\ref{acur})) which is the equivalent
 for the axial current of the Weinberg-Tomozawa
term of $\pi$-N scattering. Similarly the vector current acting on the
nucleon can create one (or more) pion via the Kroll-Ruderman term, {\it i.e.}
the contact piece of 
photoproduction (last term of eq.~(\ref{vcur})).

Let us now turn to the case of a hadronic medium. 
The expressions~(\ref{acur}) and (\ref{vcur}) illustrate in a striking fashion
the way in which the axial and vector current mixing occurs.
Indeed, in the heat bath any of the pions can be a thermal one. As an example,
 consider the Kroll-Ruderman
 term of the vector current, ignoring at this stage the denominator. 
The creation or annihilation of a thermal pion of momentum $q$ in this term
 takes care of the pion field, leaving a factor $e^{\pm iqx}$ and we are left
with a current of opposite parity, to be taken at the momentum transfer $k\pm q$
where $k$ is the photon momentum, as in the formalism of ref.~\cite{ZAH}.
 Similarly the pion production or annihilation by the Weinberg term of
the axial current introduces the vector current nuclear matrix element.  
To the extent that the Weinberg term is mediated by the rho meson
and the Kroll-Ruderman one by the $A_1$ meson, these expressions include the
effects, at low momenta, of the $\rho-A_1$ mixing. 

It is interesting to observe on expressions~(\ref{acur}) and (\ref{vcur}) that
the Kroll-Ruderman term itself can be obtained from the fourth term of the
axial current by suppression of one of the pion fields (representing creation
or annihilation of a thermal pion). Thus the three terms containing $g_A$ in
eqs~(\ref{acur}) and (\ref{vcur}) are linked together by suppression or
addition of one pion field. The same is true for the three purely pionic terms
and for the three terms in $\gamma_{\mu}$ as well. Thus a grouping three by
 three of the various terms naturally emerges from our expressions.
  
In the nuclear medium the virtual pions can be seen as a pion bath and
similar considerations about the mixing might apply. However the pions do not
come from an external reservoir but fully belong to the nucleus. Strictly
speaking there is no mixing. However the mixing terms of the currents can pick
a pion from the cloud of a nucleon introducing a similarity with the heat bath
as displayed in fig.~\ref{fig-figc} in the case of the Kroll-Ruderman term.
 The corresponding process is the
excitation of high lying nuclear states (2p-2h). In the case of the third
isospin component of the current, it is part of the well known
quasi-deuteron photoabsorption cross-section. Another
example is the influence of the Weinberg-Tomozawa term on the time part of
the axial current, which enters via the Pauli correlations and sizeably
increases the time-like axial coupling constant~\cite{KDR}. The present 
approach puts these effects, where the mixing terms of the currents pick a pion 
from the cloud, in a perspective linked to chiral symmetry.

The mixing goes along with a renormalization of certain coupling constants,
 such as the axial one, that will now be discussed.
 
\bigskip
\section{The pion decay constant}
 We start with the case of the hot pion gas. This is meant as an illustration
of our method as no new result is reached. It serves to introduce quantities
such as the residue $\gamma$ ({\it i.e.}, the wave function renormalization) 
that will be used later. 
The production of a pion by the axial current is governed by the first two
terms of the expression~(\ref{acur}). Limiting the expansion to first order in
the squared pion field we obtain: 
\begin{equation}
{\hbox{\boldmath${\cal A}$\unboldmath}}_{\mu} = 
 f_{\pi}\partial_{\mu}{\hbox{\boldmath$\phi$\unboldmath}}\;(1 -
\frac{7}{12}\frac{{\hbox{\boldmath$\phi$\unboldmath}}^2}{f_{\pi}^2}) \; .
\end{equation}
The pion field is expanded in terms of creation and annihilation operators
$B$ and $B^{\dag}$ for a quasi-pion in the medium:
\begin{equation}
{\hbox{\boldmath$\phi$\unboldmath}}(x) =
\gamma^{1/2}\int\frac{d{\hbox{\boldmath$k$\unboldmath}}}{(2\pi)^3}
\frac{1}{(2\omega_k^*)^{1/2}}
\big( {\hbox{\boldmath$B_k$\unboldmath}} +
 {\hbox{\boldmath$B_{-k}$\unboldmath}}^{\dag}\big)
 e^{i({\hbox{\boldmath$k\cdot x$\unboldmath}}-\omega_k^* t)}\; ,
\label{pifi} 
\end{equation}
where $\omega_k^*$  is the energy of a quasi-pion of momentum {\bf k},
$\omega_k^* = \sqrt{\displaystyle {\hbox{\boldmath$k$\unboldmath}}^2 
+ m_{\pi}^{*2}}$ with $m_{\pi}^*$ the effective pion mass.
 The quantity $\gamma$ is the residue of the pion pole. Since the derivative of
the pion field gives no contribution when it acts on the pions of the bath, 
the matrix element for production of a quasi-pion by the axial current reduces
to:    
\begin{equation}
\langle 0\vert{\hbox{\boldmath${\cal A}$\unboldmath}}_{\mu}(0)\vert
 \tilde{\pi}\rangle
=\frac{\gamma^{1/2}}{(2\omega_k^*)^{1/2}} if_{\pi} k_{\mu}\big(1
-\frac{7}{12}\langle\displaystyle\frac{{\hbox{\boldmath$\phi$\unboldmath}}^2}
{f_{\pi}^2}\rangle\big) = \frac{1}{(2\omega_k^*)^{1/2}} if_{\pi}^* k_{\mu}\;
\end{equation}
where the second equation defines the renormalized pion decay constant
$f_{\pi}^*$.
  In a pion gas the residue $\gamma$ has been derived by Chanfray et
al.~\cite{CEW}. To first order in the quantity 
${\hbox{\boldmath$\phi$\unboldmath}}^2$, equivalently
the pion density, it writes, in the Weinberg representation: 
\begin{equation} 
\gamma = \big(1 -
\frac{1}{2}\langle\frac{{\hbox{\boldmath$\phi$\unboldmath}}^2}
{f_{\pi}^2}\rangle\big)^{-1}\; .
\label{res}
\end{equation}
  The renormalized pion decay constant then reads:
\begin{equation}
f_{\pi}^* = f_{\pi} \gamma^{1/2}\big(1-\frac{7}{12}\langle\displaystyle
\frac{{\hbox{\boldmath$\phi$\unboldmath}}^2}{f_{\pi}^2}\rangle\big)
\approx f_{\pi} \big(1-\frac{1}{3}\langle\displaystyle
\frac{{\hbox{\boldmath$\phi$\unboldmath}}^2}{f_{\pi}^2}\rangle\big) \; .
\label{fpist}
\end{equation}
On the other hand, the temperature evolution of the condensate in a hot pion
gas is, to first order in the quantity ${\hbox{\boldmath$\phi$\unboldmath}}^2$,
 as given in ref.~\cite{CEW}:
\begin{equation}
\frac{<\overline{q} q>_T}{<\overline{q} q>_0} = 1 - \frac{1}{2}\langle
\displaystyle\frac{{\hbox{\boldmath$\phi$\unboldmath}}^2}
{f_{\pi}^2}\rangle_T \;.
\label{condT}
\end{equation}
Thus to first order in $\langle{\hbox{\boldmath$\phi$\unboldmath}}^2\rangle$ 
the renormalization of $f_\pi$ follows the evolution of the
condensate but with the coefficient 2/3, in agreement with chiral perturbation
results and other works~\cite{IOF,CEW,GL,GeL}. Note that this renormalization
 applies to both space and time components of the axial current. This agrees
with the findings of ref~\cite{EEK} where it is shown that to order $T^2$ 
Lorentz invariance is preserved.

 We now turn to the dense medium. Formally we can follow the same
procedure. The presence in the nuclear medium of a pion scalar density, in the
form of an expectation value of the quantity 
${\hbox{\boldmath$\phi$\unboldmath}}^2$, renormalizes the pion
decay constant. Formally the expression is the same as previously, 
$f_{\pi}^* = f_{\pi} \gamma^{1/2}\big(1-\frac{7}{12}\langle\displaystyle
\frac{{\hbox{\boldmath$\phi$\unboldmath}}^2}{f_{\pi}^2}\rangle\big)$.
If we treat the nuclear medium as a pion gas, the residue $\gamma$ entirely
arises from $\pi$-$\pi$ interactions and is the same as given previously in
eq.~(\ref{res}). In this simplified treatment $f_{\pi}^*$ is given by 
eq.~(\ref{fpist}). It is linked to the pion scalar density, {\it i.e.}
 the expectation value ${\hbox{\boldmath$\phi$\unboldmath}}^2$.
 Even with the simple form~(\ref{fpist}), the
renormalization of $f_{\pi}$ does not follow 2/3 of the condensate one. The
reason is that the condensate evolution in the nuclear medium is governed by
the full nucleon sigma commutator $\Sigma_N$, which is not entirely due to
${\hbox{\boldmath$\phi$\unboldmath}}^2$. There exists also the non pionic
 contribution embodied in
$\sigma_N$, as discussed previously. Thus the two renormalizations do not
follow each other. This result is general and applies as well to the axial
coupling constant $g_A$.  

This is not the only restriction which prevents a simple link to the condensate
in the nuclear medium. The residue $\gamma$ itself is not entirely due to
$\pi$-$\pi$ scattering. There exist other sources for the energy dependence of
the s-wave $\pi$-N interaction, such as the $\Delta$ excitation. The medium
renormalization of $f_{\pi}$ cannot be written in the simple 
form~(\ref{fpist}). This illustrates the complexity of the dense medium as
compared to the hot pion gas. A more phenomenological approach has been
followed by Chanfray et al.~\cite{CEK} who linked the 
in-medium pion decay constant through the nuclear
Gell-Mann-Oakes-Renner relation to the evolution of the pion mass, itself
obtained empirically from the s-wave pion-nucleus optical potential. 

\bigskip
\section{The axial coupling constant}
We now turn to the axial coupling constant. Its renormalization
 is governed by the fourth term of eq.~(\ref{acur}).
After rearrangement with the Gamow-Teller current (third term), we get:
\begin{equation}
\frac{1}{2}g_A\overline{\psi}\gamma_{\mu}\gamma_5\big({\hbox{\boldmath$\tau$
\unboldmath}} +
\frac{1}{2f_{\pi}^2}\frac{{\hbox{\boldmath$\phi\tau \cdot\phi -\tau\phi^2$
\unboldmath}}}
{\displaystyle 1 + {\hbox{\boldmath$\phi$\unboldmath}}^2/4f_{\pi}^2}\big)\psi
=\frac{1}{2}g_A\overline{\psi}\gamma_{\mu}\gamma_5{\hbox{\boldmath$\tau$
\unboldmath}}\psi\big(1-\frac{1}{3}
\langle\frac{{\hbox{\boldmath$\phi$\unboldmath}}^2/f_{\pi}^2}
{\displaystyle 1 + {\hbox{\boldmath$\phi$\unboldmath}}^2/4f_{\pi}^2}
\rangle_T\big)\; ,
\label{axialr}
\end{equation}
where on the right hand side the average is taken over the heat bath.
 On the other hand the condensate is obtained from the chiral symmetry breaking
Lagrangian ${\cal L}_{sb} = -\frac{1}{2}m_{\pi}^2{\hbox{\boldmath$
\phi $\unboldmath}}^2/( 1 + {\hbox{\boldmath$\phi $\unboldmath}}
 ^2/4f_{\pi}^2)$.
 Therefore the condensate evolution follows~\cite{CEW}:
\begin{equation}
\frac{<\overline{q} q>_{T,\rho}}{<\overline{q} q>_0} - 1 = 
-\frac{1}{2}\langle\frac{\displaystyle{\hbox{\boldmath$
\phi $\unboldmath}}^2/f_{\pi}^2}{\displaystyle 1 + {\hbox{\boldmath$\phi $\unboldmath}}
 ^2/4f_{\pi}^2}\rangle_{T,\rho} \; . 
\label{condF}
\end{equation}
 Hence the axial coupling constant renormalized by
 the pion loops (fig.~\ref{fig-figb}a) can be written: 
\begin{equation}
g_A^*/g_A = \big(1 - \frac{2}{3}\frac{<\overline{q} q>_{T}}
{<\overline{q} q>_0}\big) \; .
\label{gar}
\end{equation}
Thus with this chiral Lagrangian, in a hot medium the axial coupling constant
follows, to all orders in the pion density, 2/3 of
the quark condensate evolution (as long as it is pion dominated).
 The factor 2/3 is easily understood here: only two charges out of three
 contribute to the renormalization while all three charge
states participate in the condensate evolution. The quenching of $g_A$ is in
agreement with the universal behaviour of ref.~\cite{IOF} and with the former
result of ref.~\cite{EK}.
 We have checked the expected independence
 of our results on the particular representation of the non-linear Lagrangian.

We now turn to the case of finite density. The starting expression is the same
as the left hand side of eq.~(\ref{axialr}). In the nuclear medium the pions
 originate from the other nucleons so that the nucleon-nucleon correlations
 cannot be ignored. Here it is useful to make the link between this
renormalization and the traditional picture of meson exchange currents. We keep
only the two-body terms which are the dominant ones and work to lowest order in
the pion field.  
The corresponding graph is that of fig.~\ref{fig-figb}b. This type of exchange
graph with two pions is not usually considered in nuclear physics. It is
dictated to us only by these chiral symmetry considerations.  
We have to express the triangular graph of the figure as an effective two-body
 operator to be
evaluated between correlated two-nucleon wave functions. A simplification
occurs in the static approximation where the pions do not transfer energy to
the nucleon line. We are left with an integral over the squared pion
propagator with leads to a simple form in $x$-space for the two-body operator:

\begin{equation}                                                    
 O_{12} = - \frac{1}{6f_{\pi}^2} g_A (\gamma_{\mu} \gamma_5)_1
{\hbox{\boldmath$ \big(\tau_1 -\frac{i}{2}(\tau_1\times\tau_2)\big)
$\unboldmath}}
\varphi^2{\hbox{\boldmath$ (x_1,x_2)$\unboldmath}}\; ,  
\label{Otb}
\end{equation}
where $\varphi{\hbox{\boldmath$(x_1,x_2)$\unboldmath}}$
 is the Yukawa field, taken at the point 
\boldmath$x_1$\unboldmath, emitted by the nucleon located at the point 
\boldmath$x_2$\unboldmath\ and we have made explicit the dependence in
the isospin operator \boldmath$\tau_2$\unboldmath\ of the emitting nucleon.
 The operator $O_{12}$ has a direct and an exchange contribution. The latter
contribution vanishes for the second piece of the two-body operator
 in the limit of zero momentum current. We will furthermore ignore the exchange
term of the first piece ({\it i.e.} in \boldmath$\tau_1$\unboldmath)
 and consider only the short range correlations. We now focus on the direct
 terms and specialize to the charged currents. The isospin factors
 in eq.~(\ref{Otb}) reduce to the
 expression $3\tau_1^{\pm} - 2\tau_1^{\pm}\tau_2^{\pm}\tau_2^{\mp} =
 2\tau_1^{\pm}(1\mp\tau_2^0/2)$. The resulting contributions depend on the
relative number of protons and neutrons. In symmetric nuclear matter where they
are equal, the factor, once summed over all the pion emitters (the
nucleons with index 2), gives $2\tau_1^{\pm}$ multiplied by the nuclear density
$\rho$. In the neutron gas instead, depending whether we consider neutron decay
(the $+$ component) or proton decay (the $-$ one), we would get a factor
 $3\tau_1^+$ or $1\tau_1^-$ multiplied by the neutron density $\rho_n$.
We can summarize these results by introducing an effective density, which
depends on the charge of the current and on the neutron excess number:
\begin{equation}
       \rho_{eff}^+ = \frac{3N+Z}{2A}\rho \qquad 
\rho_{eff}^- = \frac{3Z+N}{2A}\rho \, 
\label{rhoe}
\end{equation}  
from which we recover the previous results.
Sandwiching the whole
operator $O_{12}$ between two-nucleon wave functions, we thus obtain:
  
\begin{equation}
\delta g_A^{ex}/g_A = -\frac{1}{3f_{\pi}^2}\int
d{\hbox{\boldmath$x_2$\unboldmath}}\rho_{eff}^{\pm}(x_2)[1+
G({\hbox{\boldmath$x_1,x_2$
\unboldmath}})] \varphi^2
({\hbox{\boldmath$x_1,x_2$\unboldmath}})\; ,
\label{gad}
\end{equation}
 where $ G({\hbox{\boldmath$x_1,x_2$\unboldmath}})$ is the short range
 nucleon-nucleon
 correlation function. In symmetric nuclear matter $\rho_{eff} = \rho$ whereas
a similar formula would hold in the neutron gas, with the obvious replacement
 of $\rho$ by the neutron density $\rho_n$ and of the factor 1/3 in front by 
 1/2 and 1/6 for neutron and proton decay respectively.   
 As is apparent on the expression~(\ref{gad}) it is not
 the full pion field squared which acts in the renormalization
of $g_A$, but only the part which extends beyond the range of the correlation
hole. No such distinction occurred for the pion decay constant since the pion
  produced by the axial current can be anywhere in the nucleus. Thus the
universality of the quenching which exists in the heat bath is lost.  

In order to obtain an estimate for $g_A^*$ in symmetric matter, we assume a
 total exclusion of
other nucleons in a sphere of radius $r_0 = 0.6 fm$ . In
order to facilitate the comparison of the quenching effect of $g_A$ to
that of the condensate which is governed by the nucleon sigma term, we
introduce a quantity $(\Sigma_N)_{eff}$:
\begin{equation}
(\Sigma_N)_{eff} = \frac{1}{2}m_\pi^2\int d{\hbox{\boldmath$x$\unboldmath}}
\theta(x-r_0) {\hbox{\boldmath$\varphi^2(x)$\unboldmath}} \; .
\label{sigeff}
\end{equation}
Numerically for point-like pion emitters, we find an effective value
 $(\Sigma_N)_{eff}\approx 21 MeV$. It is interesting to compare this value
with a model calculation in the quark picture. We have used the results of
Wakamatsu~\cite{WA} in a chiral soliton model.
The quantity
${\hbox{\boldmath$\varphi$\unboldmath}}^2({\hbox{\boldmath$x$\unboldmath}})$ 
is replaced by the sea quark density distribution according to:
$\frac{1}{2}m_\pi^2 {\hbox{\boldmath$\varphi^2(x)$\unboldmath}} \to
     2m_q\overline{q} q{\hbox{\boldmath$(x)$\unboldmath}} $.
This gives a very similar value $(\Sigma_N)_{eff}\approx 19 MeV$.
 
However these numbers 
do not include the Pauli blocking effect which removes
the occupied states in the process of pion emission. This effect
has been calculated in refs.~\cite{ERC,CE} but for the whole space integral
({\it i.e.} without a cut-off) of the quantity
 ${\hbox{\boldmath$\phi$\unboldmath}}^2$.
 Expressed in terms of a modification of the sigma commutator it amounts
to a reduction $(\Delta\Sigma_N)_{Pauli} = -2.6 MeV $.
 The blocking effect, which is moderate, should 
 be even less pronounced with the cut-off. We ignore it in the following. 

Coming back to the renormalized axial coupling constant, we have:
\begin{equation}
g_A^*/g_A = 1 - \frac{2}{3}\frac{\rho(\Sigma_N)_{eff}}{f_{\pi}^2m_{\pi}^2}\; .
\label{gaq}
\end{equation}  
This represents a 10\% quenching at normal nuclear density in symmetric matter
(15\% for neutron decay in a neutron gas of the same density),
 while the condensate
has dropped by 35\%. Notice that the evolution of $g_A$ is sizeably slower. 
This quenching applies to all the components, space or time, of the axial
current. Other renormalization effects have to be added. They are known to
 act differently on the
different components. For instance the Weinberg-Tomozawa term acts on the
time component alone, producing a sizeable enhancement~\cite{KDR}. In the case
of the space component the nucleon polarization under the influence of
the pion field $N \to \Delta$ leads to the Lorentz-Lorenz
 quenching~\cite{EFT}. In the latter case the two
renormalizations go in the same direction of a quenching. The extra reduction
 that we have introduced in this work could help to explain
 the large amount of quenching observed in
Gamow-Teller transitions. To get an idea, we fictitiously translate the
reduction by chiral symmetry into an equivalent Lorentz-Lorenz effect. We
introduce an effective Landau-Migdal parameter $\delta g_{N\Delta}'$, to be
added to the genuine one, so as to reproduce the 10\% quenching. This
corresponds to an increase $\delta g_{N\Delta}' \approx 0.16$, a significant
increase. Indeed the quenching of the Gamow-Teller sum rule requires, if all
attributed to the Lorentz-Lorenz effect, 
$g_{N\Delta}'$ to be as big as 0.6 - 0.7 while the favoured theoretical value 
 is around 0.4~\cite{DIC}. Hence the chiral induced quenching would help to fill
 the gap.

Closely related to the Gamow-Teller transition is the pion photoproduction at
threshold through the Kroll-Ruderman term. To lowest order the nuclear
transition is governed by the axial current. We want now to discuss how it is
renormalized in the medium following chiral symmetry requirements. Expanding to
first order in ${\hbox{\boldmath$\phi$\unboldmath}}^2/f_{\pi}^2$
 and applying Wick theorem, the relevant current writes:
\begin{equation}
({\hbox{\boldmath${\cal V}$\unboldmath}}_{\mu})_{KR} =
-\frac{g_A}{2f_{\pi}}\overline{\psi}\gamma_{\mu}\gamma_5
({\hbox{\boldmath$\tau\times\phi$\unboldmath}})
\psi\big(1 - \frac{5}{12}\langle\displaystyle\frac{
{\hbox{\boldmath$\phi$\unboldmath}}^2}{f_{\pi}^2}\rangle\big) \; .
\label{KR}
\end{equation}
For the production of a quasi-pion in the medium, the renormalization $r_{KR}$
 of the amplitude involves again the residue $\gamma$:
\begin{equation}
r_{KR} = \gamma^{1/2}\big(1 - \frac{5}{12}\langle\displaystyle\frac{
{\hbox{\boldmath$\phi$\unboldmath}}^2}{f_{\pi}^2}\rangle\big) \; .
\end{equation}
 
For illustrating the complexity of the situation we first assume that the
 residue is entirely given by $\pi-\pi$ scattering and
take the value of eq.~(\ref{res}). Moreover we ignore the correlation
 complications. We obtain then: 
\begin{equation}
r_{KR} = \big(1 - \frac{1}{6}\langle\displaystyle\frac{
{\hbox{\boldmath$\phi$\unboldmath}}^2}{f_{\pi}^2}\rangle\big) \; ,
\label{rKR}
\end{equation}
which is 1/3 of the variation of the condensate, in contradistinction to the
axial transitions where the factor is 2/3. This result does not contradict the
general expressions of Dey et al.~\cite{IOF} as the Kroll-Ruderman term
represents already a mixing of the axial current into the vector one. This
reduction factor could apply to other mixing amplitudes, but we have not
established it. 
The evolution as 1/3 of the condensate one would apply in the hot pion gas
situation. In the nuclear medium all the complications mentioned previously
occur: the role of the correlations, the link between the condensate evolution
and the expectation value of ${\hbox{\boldmath$\phi$\unboldmath}}^2$ and
 the problem with the residue $\gamma$.
This case cumulates all of the difficulties of the dense medium. In all
instances the overall renormalization of the Kroll-Ruderman matrix element
in the nuclear medium should be small.

\section{Conclusion}
\begin{figure}
\begin{center}
\leavevmode
\epsfxsize = 7.cm
\epsfig{file=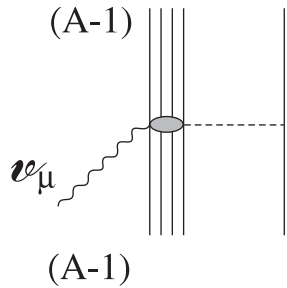}
\end{center}
\caption{\label{fig-figa}Illustration of a mixing effect in the nucleus 
by the Kroll-Ruderman term.}
\end{figure}

In conclusion we have investigated the behaviour of the nuclear medium in
relation with chiral symmetry restoration. We have
focused
on the extension of the parity mixing concept between the axial and vector
correlators, which exists in the hot pion gas. In the nuclear medium there is
 no mixing {\it stricto sensu}. Indeed the pions, which induce the 
mixing, are not part of an
external system, as in the thermal case, but they belong to the virtual pion
cloud which is an integral part of the nucleus. We have shown that nevertheless 
certain consequences of the mixing survive. The nucleus behaves in certain
respects as a pion reservoir. The virtual pion emitted by a nucleon acts, as
illustrated in fig.~\ref{fig-figa}, on the
remainder of the nucleus, {\it i.e.} on the system of (A-1) nucleons, as the
pion of the heat bath. The mixing which does not exist at the level of the
whole nucleus is present only at the sublevel of the (A-1) nucleon system. This 
translates by the fact that, in the ``mixing'' cross-sections (such as the
quasi-deuteron photoabsorption one), at least one nucleon has to be ejected: 
the emitter or absorber of the
 pion. As for the heat bath, this pion reservoir produces a quenching
of the axial coupling constants. Since the pion originates from a neighbouring
nucleon this renormalization is nothing else than a meson exchange
contribution. It involves the exchange of two pions and has not been so far 
considered,
 to our knowledge. We have expressed the renormalizations in terms of the
pion scalar density. The same quantity also enters in the quark condensate
evolution. One can therefore think of a link between the two quantities, as
occurs in the heat bath where the link is simple. There is however an important
difference of the nuclear medium with respect to the heat bath: the
renormalizations are not described by a universal quenching factor expressed in
terms of the average squared pion field. The nucleonic observables such as
$g_A$ are renormalized differently due the sensitivity to nucleon-nucleon
short range correlations. In this case only that part of the pionic field
 which is beyond
the correlation hole enters in the renormalization.
This prevents the link to the condensate evolution which instead
involves the average scalar density. This is an illustration of
the point made by T. Ericson~\cite{TE} about the possible importance of
 the spatial fluctuations of the condensate. We have
given an estimate for the quenching of the axial coupling constant arising
from the requirements of chiral symmetry. Although it is not very large (about
10\%), this additional quenching is significant and may help explain the large observed
quenching of the Gamow-Teller sum rule.
We have also discussed the photoproduction amplitude arising from the
Kroll-Ruderman term. It represents a mixing term of the axial current
into the vector one. We have shown that its evolution is slower than the
axial coupling constant one.

This work can be extended to enlarge the space. The first step is to
include the Delta excitation. Another extension concerns the explicit
introduction of the rho and the $A_1$ mesons, which the mixing of the axial and
vector correlators allows to be excited either by the vector or by the
axial current.

\bigskip
We thank Prof. T. Ericson for useful comments. We are very grateful to
 Prof. M. Wakamatsu for
communication of his detailed results on the quark scalar density.

\newpage

\begin{figure}
\begin{center}
\leavevmode
\epsfxsize = 14. cm
\epsfig{file=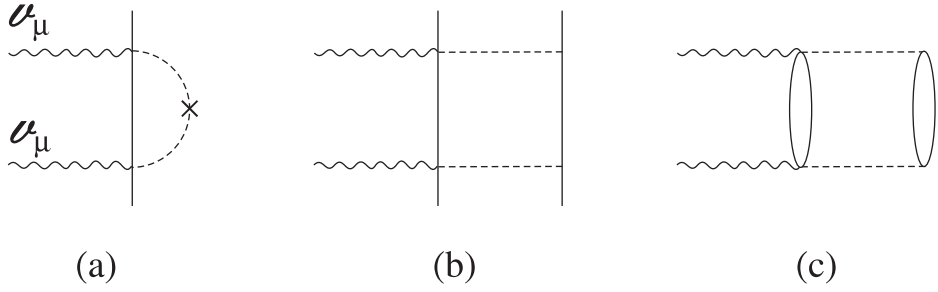}
\end{center}
\caption{\label{fig-figc}Illustration of a mixing effect in the vector
 correlator (a) in the heat bath (denoted by a cross),
 (b) equivalent diagram in the
nucleus with its translation (c) in many-body diagrams.}
\end{figure}

Figure captions: 
\vskip 4mm
Fig.\ref{fig-figc}: Illustration of a mixing effect in the vector correlator 

(a) in the heat bath (denoted by a cross), (b) equivalent diagram in the

nucleus with its translation (c) in many-body diagrams.

\bigskip

\begin{figure}
\begin{center}
\leavevmode
\epsfxsize = 14. cm
\epsfig{file=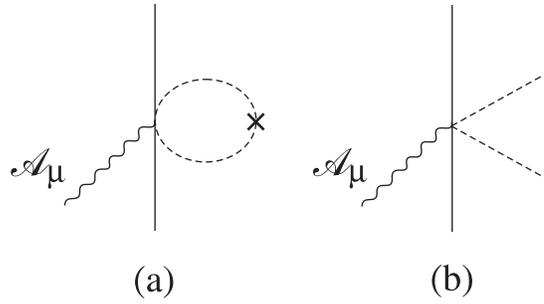}
\end{center}
\caption{\label{fig-figb}Renormalization of the nucleonic axial coupling 
constant 
(a) - by a pion loop in the hot pion gas (the cross denotes the heat bath),
(b) - by the virtual pion cloud in
 the nucleus. 
}
\end{figure}

Fig.\ref{fig-figb}: Renormalization of the nucleonic axial coupling constant
 
(a) - by a pion loop in the hot pion gas (the cross denotes the heat bath),

(b) - by the virtual pion cloud in
 the nucleus. 

\bigskip

Fig.\ref{fig-figa}: Illustration of a mixing effect in the nucleus by the
 Kroll-Ruderman term.

\newpage

\end{document}